\documentclass[twocolumn,showpacs,aps]{revtex4}
\usepackage{epsf}
\def\be{\begin{equation}}
\def\ee{\end{equation}}
\def\beq{\begin{equation}}
\def\eeq{\end{equation}}
\def\hb{\hbar}
\def\ep{\varepsilon}

\def\rtf{\rho_{TF}}
\def\r0{\rho_{0}}
\def\la{\lambda}
\def\cc{{\cal C}}
\def\tx{{\tilde x}}
\def\ty{{\tilde y}}
\def\tz{{\tilde z}}
\def\tv{{\tilde v}}

\begin{document}

\title{Dissipative flow and vortex shedding in the Painlev\'e boundary layer of a Bose 
Einstein condensate}

\author{Amandine Aftalion}
\email{aftalion@ann.jussieu.fr}
\affiliation{CNRS and Laboratoire Jacques-Louis Lions,  Universit\'e 
Paris 6, 175 rue du Chevaleret, 75013 Paris,
France.}
\author{Qiang Du}
\email{qdu@math.psu.edu}
\affiliation{Department of Mathematics, Penn State University, 
University Park, PA 16802, USA.}
\author{Yves Pomeau}
\email{pomeau@lps.ens.fr}
\affiliation{CNRS and Laboratoire de Physique Statistique, Ecole 
normale sup\'erieure, 24 rue Lhomond, 75231 Paris cedex 05, France.}
\date{\today}


\begin{abstract}
Raman et al.\cite{R} have found experimental evidence for a critical 
velocity under which there is no dissipation when  a 
detuned laser beam is moved in a Bose-Einstein condensate. We analyze the 
origin of this critical velocity in the low density region close to 
the boundary layer of the cloud. In the frame of the laser beam, we do a blow up on this low density region which can be described by a 
Painlev\'e equation and write the approximate equation satisfied by the wave function in this region. We find
   that there is always a drag around the laser beam. Though the beam passes through the surface of the cloud and
the sound velocity is  small in the Painlev\'e boundary layer, the shedding of vortices starts only when a threshold velocity is reached. This critical velocity is lower than the critical velocity computed for the corresponding 2D problem at the center of the cloud. At low velocity, there is a stationary solution without vortex and the drag is small. At the onset of vortex shedding, that is above the critical velocity, there is a drastic increase in drag.
\end{abstract}

\maketitle

Dilute  Bose-Einstein condensates have recently been achieved in
confined alkali-metal gases  and the study of vortices 
therein
is one of the key issues. Raman et al. \cite{R,RO}, Onofrio et al. 
\cite{O} have studied dissipation in a Bose Einstein condensate by 
moving a blue detuned laser beam through the condensate at different 
velocities. They found experimentally a critical velocity for the 
onset of dissipation. 
This critical velocity has been related to the one found by Frisch et al. \cite{FPR} for the problem of a 2D superfluid flow 
around an obstacle in the framework of Nonlinear Schrodinger Equation 
(NLS): below a critical velocity, the flow is stationary and dissipationless, while
  beyond this critical velocity, the flow around the 
disc becomes time dependent and vortices are emitted. 
Numerical simulations have been done for this type of problem in 2D 
\cite{HB} and 3D \cite{JMA,W}.  In particular, the direct 3D simulation of \cite{W} shows the plot of the drag against the velocity. A critical velocity can be numerically computed when the drag 
becomes nonzero, but no precise mechanism of vortex nucleation is described by the authors.
 This critical velocity has been analyzed theoretically for a homogeneous 2D system \cite{SZ} and an inhomogeneous 2D system \cite{C,FS}. 

In this paper, we want to  take into account the 3D geometry of the experiment of \cite{R,O, RO}. Our aim is to understand the mechanism of vortex nucleation in the boundary region. Indeed the analysis of \cite{FPR} allows to understand what is 
happening in the interior  of the cloud, where  the kinetic energy is negligible in front of the interaction energy. 
 In the region where the laser beam crosses the boundary of 
the cloud, the sound velocity gets  small, since the amplitude of 
the wave function becomes  small. There, the kinetic energy term can no longer be neglected in front of the trapping and interaction terms. We blow up this region in such a way that the trapping potential varies linearly with the distance to the boundary and far away from the laser beam, the wave function is then given by a Painlev\'e equation.
We analyze the behavior of the wave function in the frame of the laser beam. The real experiments are quite complex, and in particular here we do not take into account the oscillations and acceleration of the beam but we believe that our analysis allows to understand the mechanism of increase of drag. 
One of our main results is that there is always a drag around the laser beam and this drag grows continuously. At low velocity, the drag is not
a consequence of the shedding of vortices, and finally of a time
dependent density and velocity field. The origin of this drag is in
the radiation condition for the wavefield: the motion
changes continuously the structure of the solution seen in the frame
of reference of the "fluid" at infinity.
 We study the  transition  toward a time
dependent regime of vortex shedding, which happens at a critical velocity.  
  The critical velocity that we find is lower than the 2D critical velocity at the center of the cloud coming from the computation of \cite{FPR}.
 Vortices are nucleated close to the boundary of the cloud and the tubes grow and detach to form rings that move downstream.  When tubes are emitted, 
significantly large drag values are observed. The drag increases smoothly as the velocity increases.

The dynamics can be modeled using the Gross Pitaevskii equation at zero temperature with 
an external trapping potential $V_{tr}=m/2( \omega_x^2 x^2+\omega_y^2 y^2 + \omega_z^2 z^2)$.
$$i\hb\partial_t \Psi=-{{\hb^2}\over{2m}}\Delta 
\Psi+(V_{tr}+Ng|\Psi|^2)\Psi.$$
If an object is moved inside the condensate, $V_{tr}$ has to be replaced 
by $V_{tr}+V_{ob}$, where $V_{ob}$ depends on $x-vt$.
Based on the experimental data of \cite{R,O}, we take $a=mg/4\pi\hb^2=2.94\ nm$, 
$N=1.2.10^{7}$, $\omega_y=\omega_z=377s^{-1}$, 
and $\omega_x=\la\omega_z$, with $\la=0.3$. We also define the characteristic 
length  $d=(\hb / m\omega_z)^{1/2}=2.71 \ \mu m$ and a small
nondimensionnalized parameter $\ep$ given by 
$$\ep= ({{d}\over {8\pi Na}})^{2/5}.$$ 
We find that $\ep=6.21 \ 10^{-3}$ which may be
viewed as small parameter and allow rescaling the equation near the edge of the condensate. Re-scaling the distances 
by $R=d/\sqrt \ep=34.4\ \mu m$, the time by $1/(\ep\omega_z)$,  
we have $\psi({\bf r},t)=R^{3/2} \phi({\bf \tilde r},{\tilde t})$ where ${\bf 
\tilde r}=R{\bf r}$. 
In these new units,   
 the radii of the condensate are $R_y=R_z=0.65$ and $R_x=2.18$. 
The laser beam is modeled by an
obstacle which is a cylinder $\cc$ of axis $z$ and radius $l=0.19$ on which $\psi=0$.
We will work in the frame where the obstacle is stationary. Outside the obstacle, the equation can be rewritten as
$$-2i\partial_t \psi
=\Delta\psi +{1\over \ep^2}
(\rtf-|\psi|^2)\psi,$$
where
$\rtf=\r0 -(\la^2x^2+ y^2+z^2)$ is the Thomas Fermi limit density and $\r0=0.42$ is the rescaled chemical potential. Note that 
 $|\psi|^2$ is
 close to its Thomas Fermi value $\rtf$ 
except near the obstacle and near the boundary of 
the cloud. 
 This  boundary layer has a thickness  of order $\ep^{2/3}$ so that we rescale 
 the domain with $\psi(\tx,\ty,\tz)=\ep^{1/3}u(x,y,z)$, where 
$x=\tx/\ep^{2/3}$, $y=\ty/\ep^{2/3}$ and 
$z=(\sqrt{\r0}-\tz)/\ep^{2/3}$, $v=\tv \ep^{2/3}$. By blowing up 
the boundary of the cloud near $z=0$, and truncating at $z=L$,
the rescaled layer thickness,  we see that 
the modulus of the stationary solution in the boundary layer
for $|x|$ and $|y|$ large,  that is far away from the obstacle, 
is given by the solution of the first Painlev\'e equation \cite{FF,DPS}.
\beq\label{pain}
p''+(2z\sqrt{\r0}-p^2)p=0,\ p(0)=0,\ p(L)=\sqrt{2\sqrt{\r0}L}.
\eeq
We choose the size of the boundary layer $L$ so that 
$\ep^{2/3}L=3\sqrt{\r0}/10$. This is based on the consideration that, on the one hand, 
$\ep^{2/3}L$ should be suitably small so that  
$2z\sqrt{\r0}$ is a good 
approximation for $\rtf=\r0 -\tz^2$ in the boundary layer
and on the other hand the critical velocity at $z=L$ is not 
too different from the critical velocity at the center of the cloud. 
The obstacle is now a cylinder of radius $a=l/\ep^{2/3}=5.6$.

The obstacle moves at the rescaled velocity $v=v_{exp}/(\ep^{1/3}\omega_z R)$, and in the  frame of  the obstacle, 
 the equation becomes
\begin{equation}\label{equ}
-2i\partial_t u= 
\Delta u-2 iv\partial_x u+(2z\sqrt{\r0}-|u|^2)u.
\end{equation}

We want to understand the behaviour of solutions depending on $v$. If we restrict (\ref{equ}) to $z=L$, we can perform a similar analysis to \cite{FPR} and get the value of the critical velocity for the onset of vortex shedding and find
 $v_c^2=2\sqrt{\r0}L/11=2c_s^2/11$, where $c_s$ is the sound velocity. Of course, we cannot apply this analysis in the low density region, since there the sound velocity gets close to 0. Another mechanism has to be understood.
The rescaled drag around the obstacle is
\beq\label{drag} drag =\frac{1}{2}\int_{\cal C} (u_x \bar{u}_n-\bar{u}_x u_n) \; dl\ dz.\eeq

\begin{figure}[htb]
\centerline{\epsfxsize=3.in\epsfbox{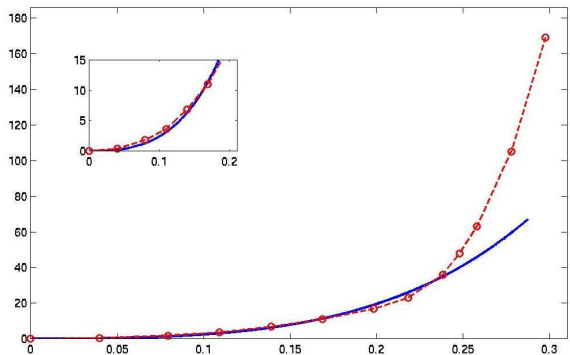}}
\vspace{-0.1in}
\caption{Drag vs. $v/c_s$: 
$--$ for (\ref{d2}), $-o-$ for numerical solution of (\ref{equ});
insert: zoomed in for small $v$.}
\vspace{-0.1in}
\label{fig1}
\end{figure}

We first analyze the stationary solution of (\ref{equ})  
in the very low density region, where
the system is very dilute and one can neglect the nonlinear term.
In fact, a precise condition is that $p^2$ is less than $v^2$, which gives a truncation point $z_c$ at which $p^2(z_c)=v^2$.
It is rather straightforward in classical scattering theory to 
compute the perturbed wavefield and finally the drag on the 
obstacle (a related problem, the scattering of sound by a 
cylinder, is treated in \cite{M}).
In the low density region, it is reasonable to look for $u$ with the following ansatz
\beq\label{sol}u(x,y,z)=p(z)\psi(x,y)e^{ivx}.\eeq
We can first approximate $p(z)$ in this region by an Airy function given by 
the solution of $p''+2zp\sqrt{\r0}=0$, 
that is, by defining $\overline{z}^3=1/(2\sqrt{\r0})$, we have
\beq\label{airy}
p(z)\approx \sqrt{2} Ai({{-z}\over{\overline z}})\approx
{1\over {\sqrt{2\pi}}} ({{{-\overline{z}}\over{ z}}})^{1\over 4}exp(-{2\over 3}({{-z}\over{\overline z}})^{3\over2}).
\eeq
Then, outside the obstacle,
 $\psi$ is a solution of the 2D Helmholtz equation
\beq\label{helm}
\Delta \psi+v^2\psi=0   
\eeq
with $\psi=0$ for $r=a$, the obstacle boundary,
 and $\psi\approx e^{ivx}$ at infinity.
This solution can be computed \cite{M} in terms of Bessel functions  $J_k$ and $N_k$.
One finds that, at  leading order  for $v$ small, the 2D drag  of $\psi$ is proportional to
\beq\label{d1}
v^2J_0^2(v)J_1(v)N_2(v)/N_0^2(v) \sim v/\log^2 v.\eeq
The total drag has to be multiplied by the integral of $p^2$ along the $z$ axis
to the  truncation point $z_c$ defined by $p^2(z_c)=v^2$. Direct
calculation gives
\beq\label{d2}
{v\over{\log^2 v}}\int_{-\infty}^{z_c}  p^2\ dz\approx C{{v^3}\over {\log^{8/3} v}}.\eeq
In conclusion, the total scattering drag  tends to zero at low speed. It is plotted in Figure \ref{fig1} (solid line).

To understand how the
solutions of (\ref{equ}) behave, we
numerically integrate the equation
in a computational domain of
dimension $60\times 60 \times L$ with
periodic boundary conditions in $x$ and $y$ and taking $u=0$ 
on the boundary of the obstacle and away from the condensate ($z=0$). 
At the truncated surface $z=L$ inside the condensate, we use
the condition $({\partial}/{\partial z})(u/p) =0$,
where  $p$ is the solution to the Painlev\'e equation (\ref{pain})
described before.
 The numerical solution is computed based on a continuous piecewise quadratic 
finite element approximation in space and the Runge-Kutta fourth-order
in time integration scheme.
Using $p$ as   initial condition, we first compute the solution 
of (\ref{equ}) for 
some time by adding a damping coefficient, that is, 
we replace $iu_t$  in (\ref{equ}) by $iu_t(1+i \gamma )$.
For small velocity, this effectively drives the numerical solution of
(\ref{equ}) close to a stationary solution.
Then, we continue the 
integration with a much reduced damping coefficient $\gamma = 0.02$ or with no damping $\gamma =0$.

In what follows, we will divide the 
 velocity 
 by the sound velocity  at center
 $c_s=\sqrt{2\r0}/\ep^{1/3}$.  In Figure \ref{fig1}, we plot the drag vs.
the velocity divided by $c_s$. For a given velocity value, the drag 
 is the value obtained through time averaging of (\ref{drag}).
  We have verified that with different 
small values of $\gamma$, the drag calculation remains essentially the same.

\begin{figure}[htb]
\centerline{
\hspace{-0.1in}
\epsfxsize=1.6in\epsfbox{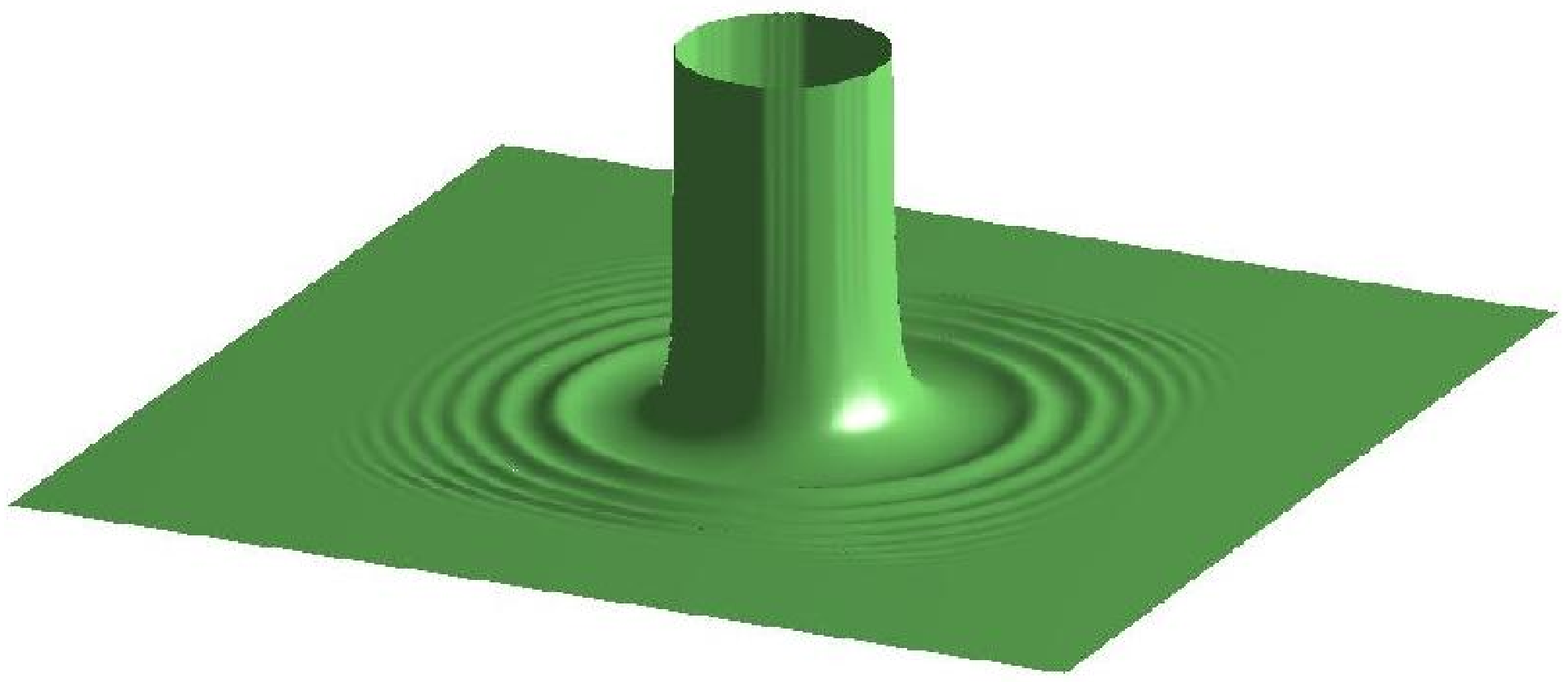}
\epsfxsize=1.6in\epsfbox{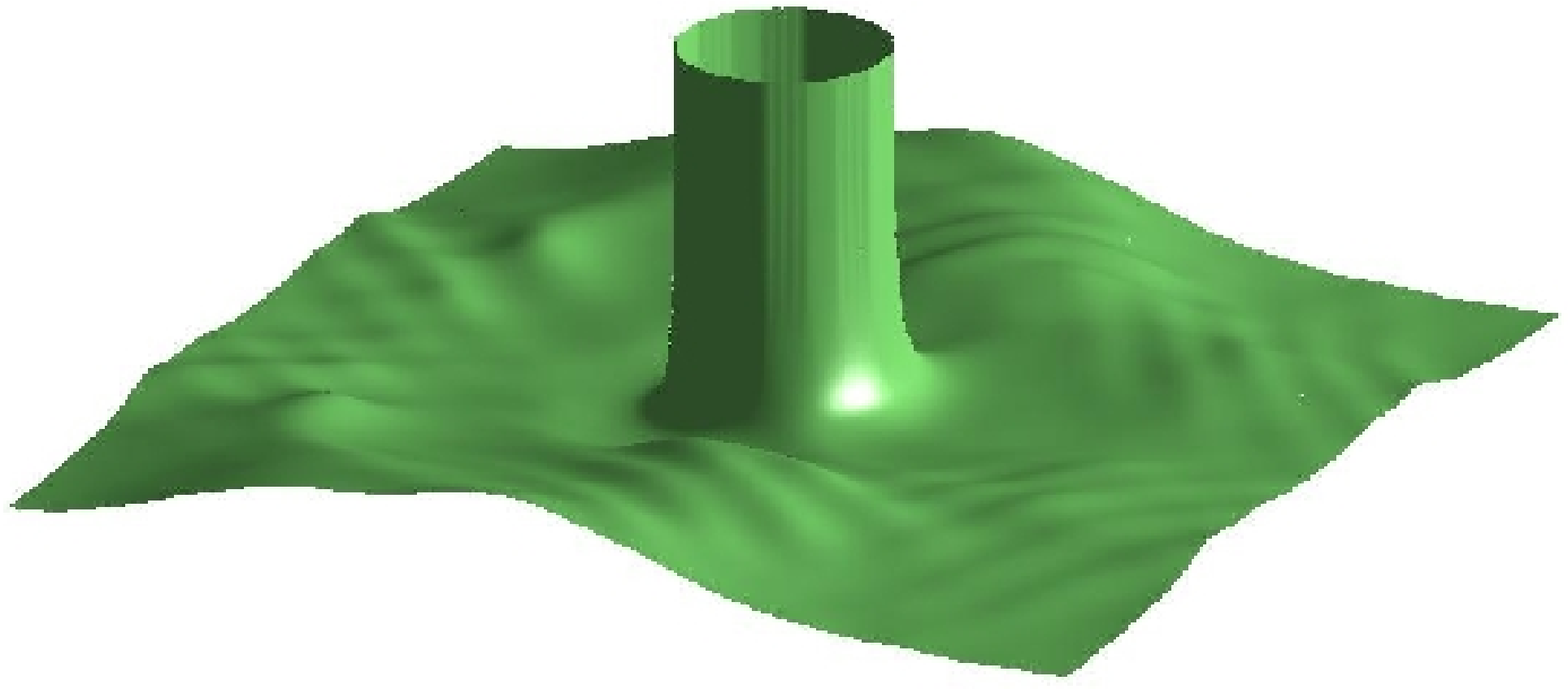}}
\vspace{-0.1in}
\caption{Isosurface 
snapshot of $|u|$: surface waves for $v=0.08$ and $v=0.2$.}
\label{fig2}
\vspace{-0.1in}
\end{figure}

For $v$ small, we find that the solution is almost stationary. 
 Surface oscillations are present
 near $z=0$, and the drag is small, but not zero. See Figure 
\ref{fig2} for plots of the solution. 
 The drag computed in this regime fits very well to the cubic growth 
given by (\ref{d2}).

\begin{figure}[htb]
\centerline{\hspace{-0.1in}
\epsfxsize=1.6in\epsfbox{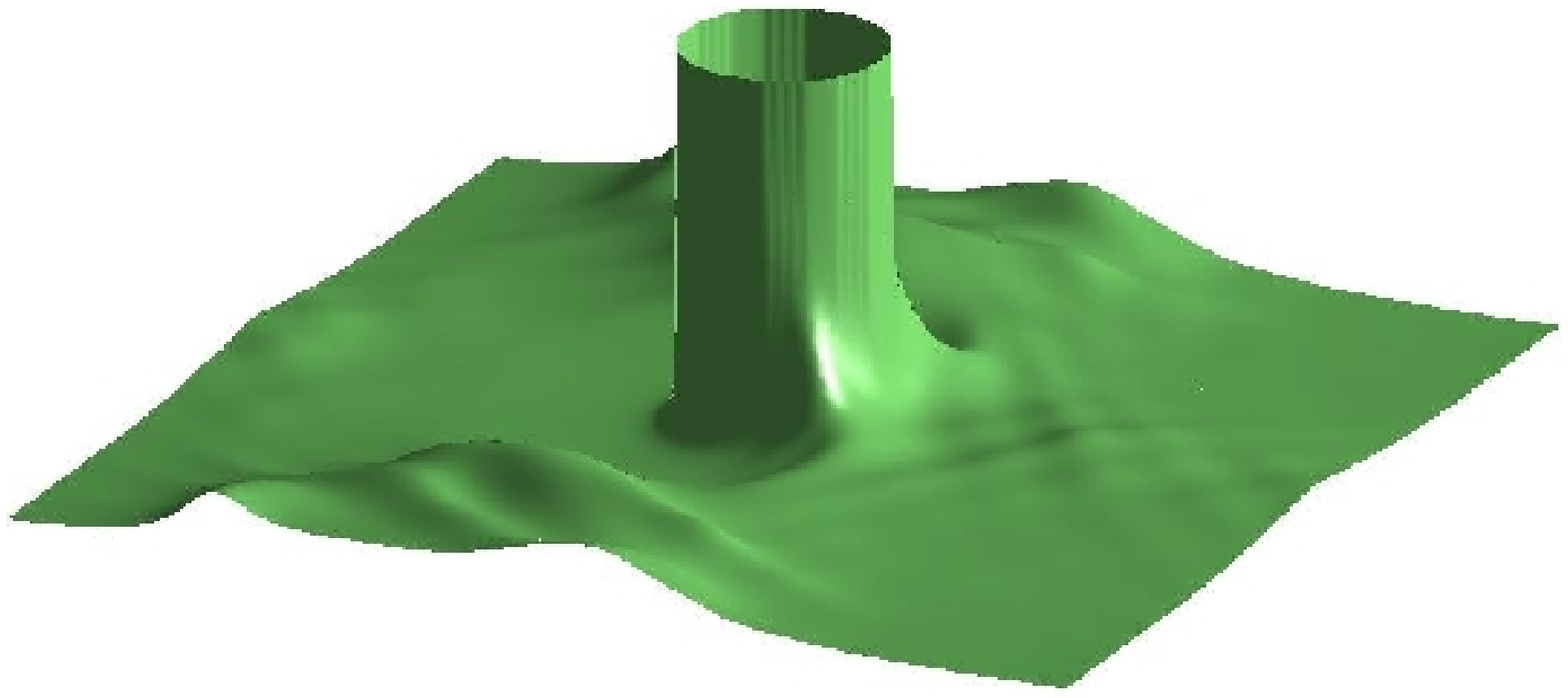}
\epsfxsize=1.64in\epsfbox{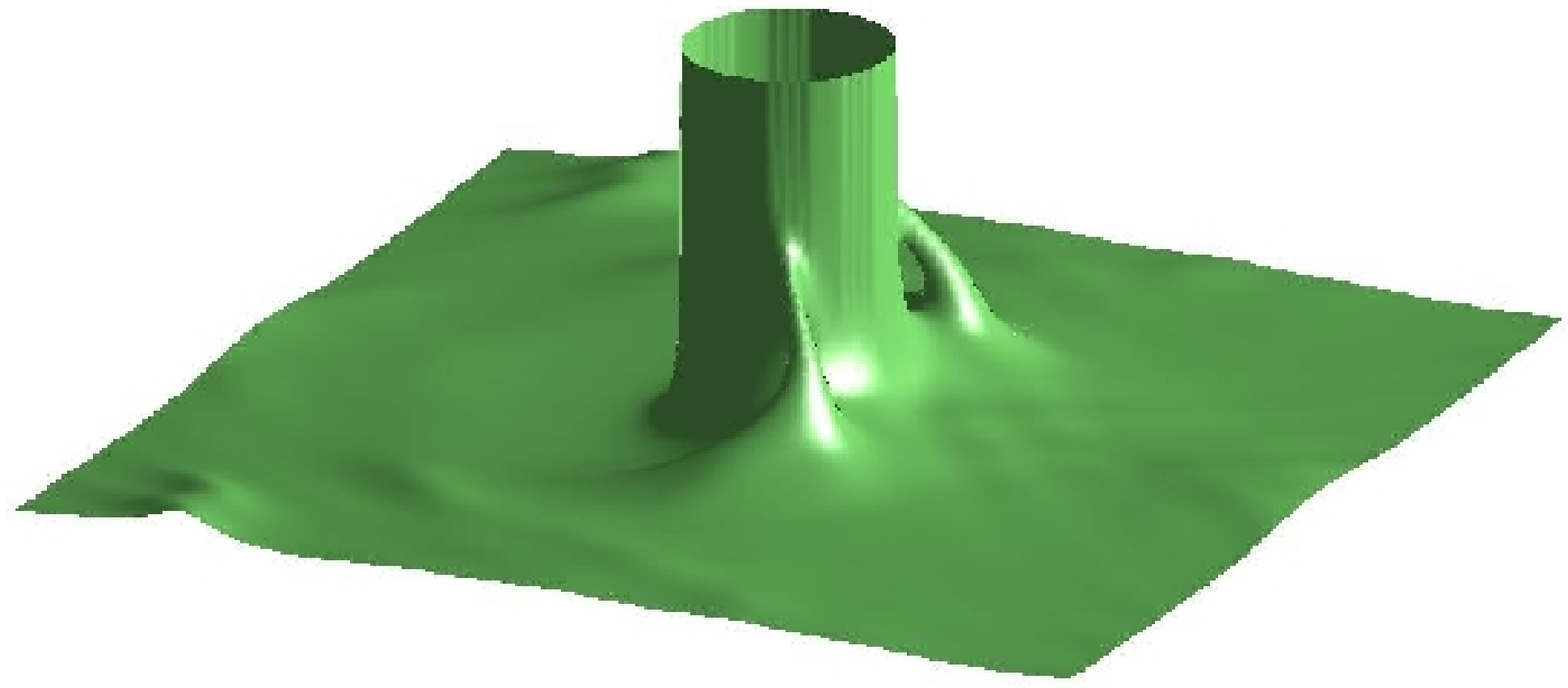}}
\vspace{-0.1in}
\caption{Isosurface 
snapshots of $|u|$ at different times for $v=0.24$: formation of vortex handles.}
\vspace{-0.1in}
\label{fig3}
\end{figure}

When $v$ is increased, at a critical velocity $v_c/c_s\approx 0.2$, the surface oscillations develop into small handles that move up and down the obstacle without detaching; see Figure \ref{fig3}.

 There is no stationary solution, but no vortex shedding either: the small handles move up the obstacle to a critical $z$ value and down.  This instability may be related to the one discussed by Anglin \cite{A1}: in our scaling, the critical velocity found in \cite{A1} is 0.2. At this stage, the solutions do not produce large drag nor vortex shedding. 

It is only for larger velocities ($v/c_s >0.25$) that the handles move up to the top, detach from the obstacle and produce significant drag. This is a wholly nonlinear phenomenon and most likely cannot be described by a linear analysis.

\begin{figure}[htb]
\centerline{
\epsfxsize=2.6 in\epsfbox{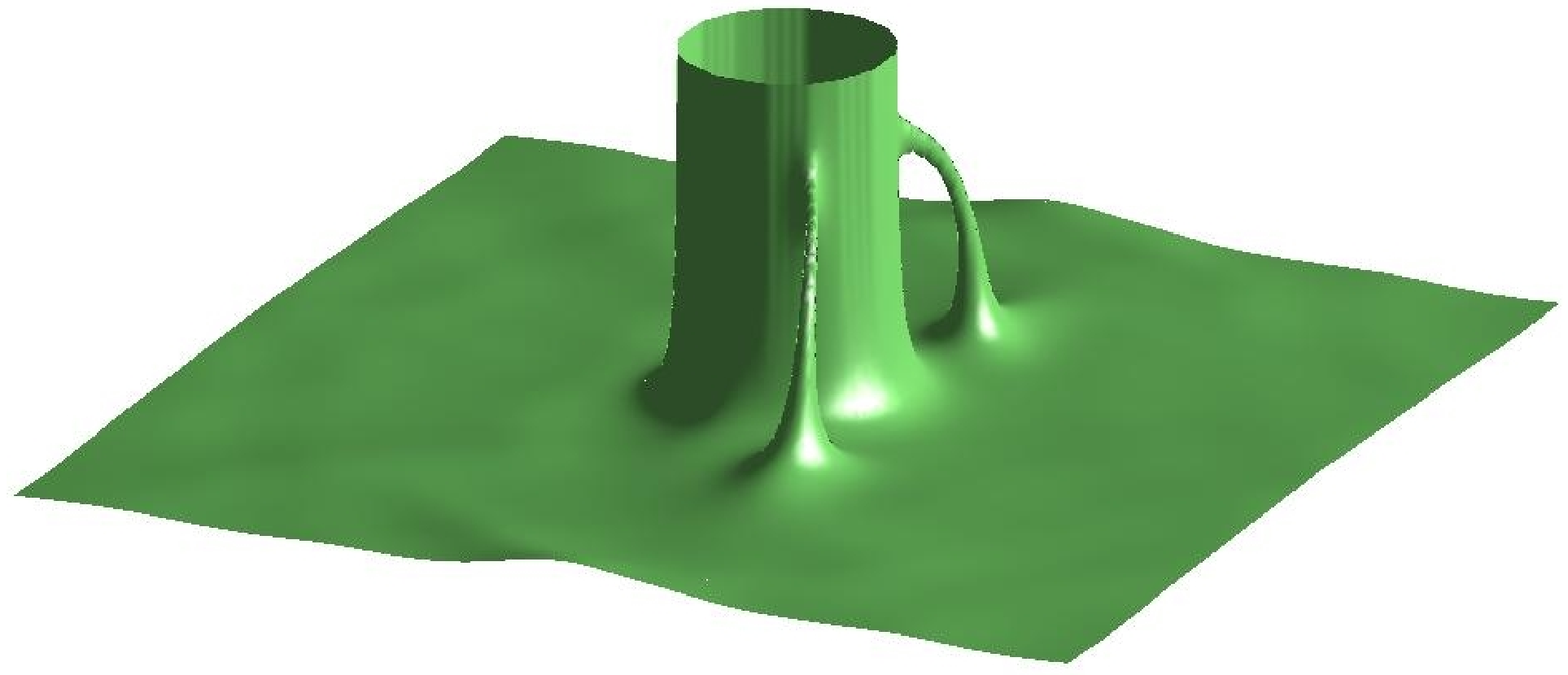}
$\;$
}
\centerline{
         \epsfxsize=2.6 in\epsfbox{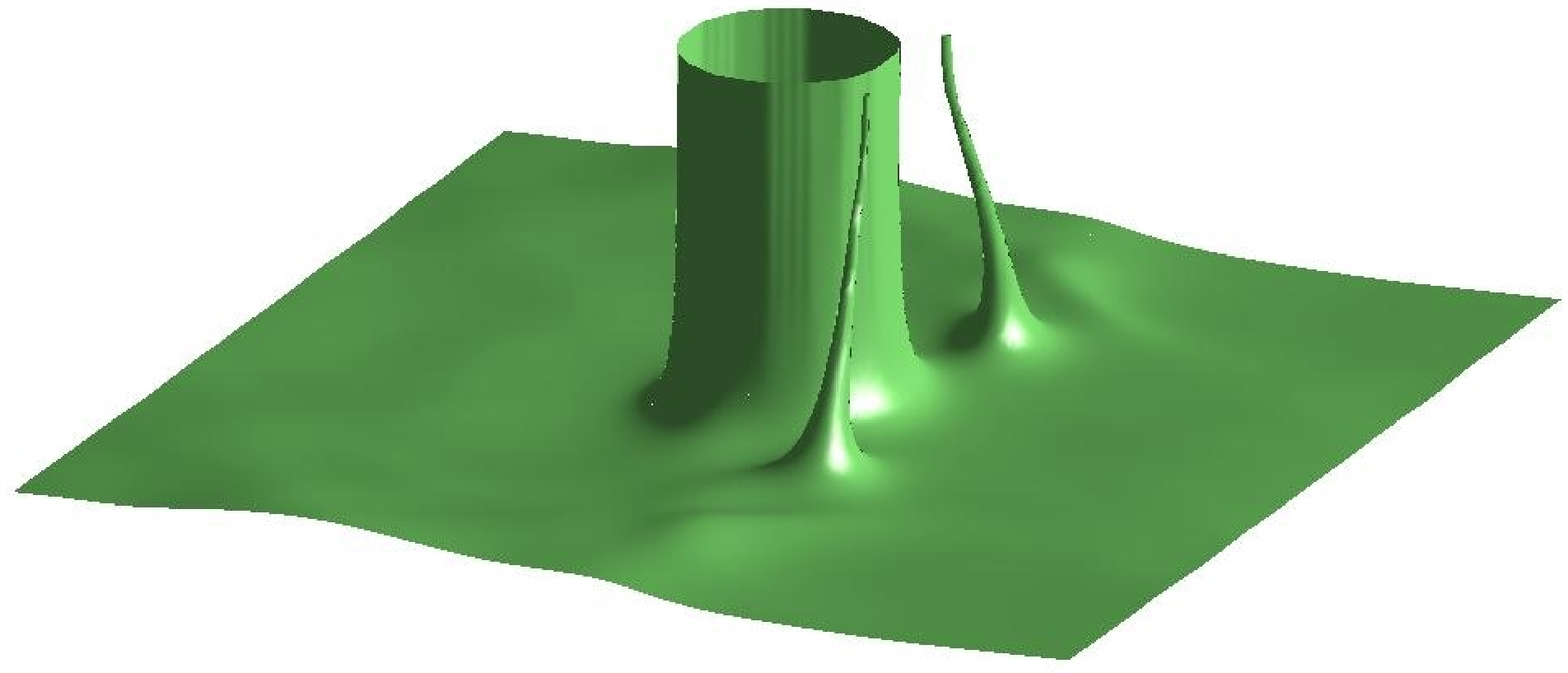}}
\centerline{\epsfxsize=2.6 in\epsfbox{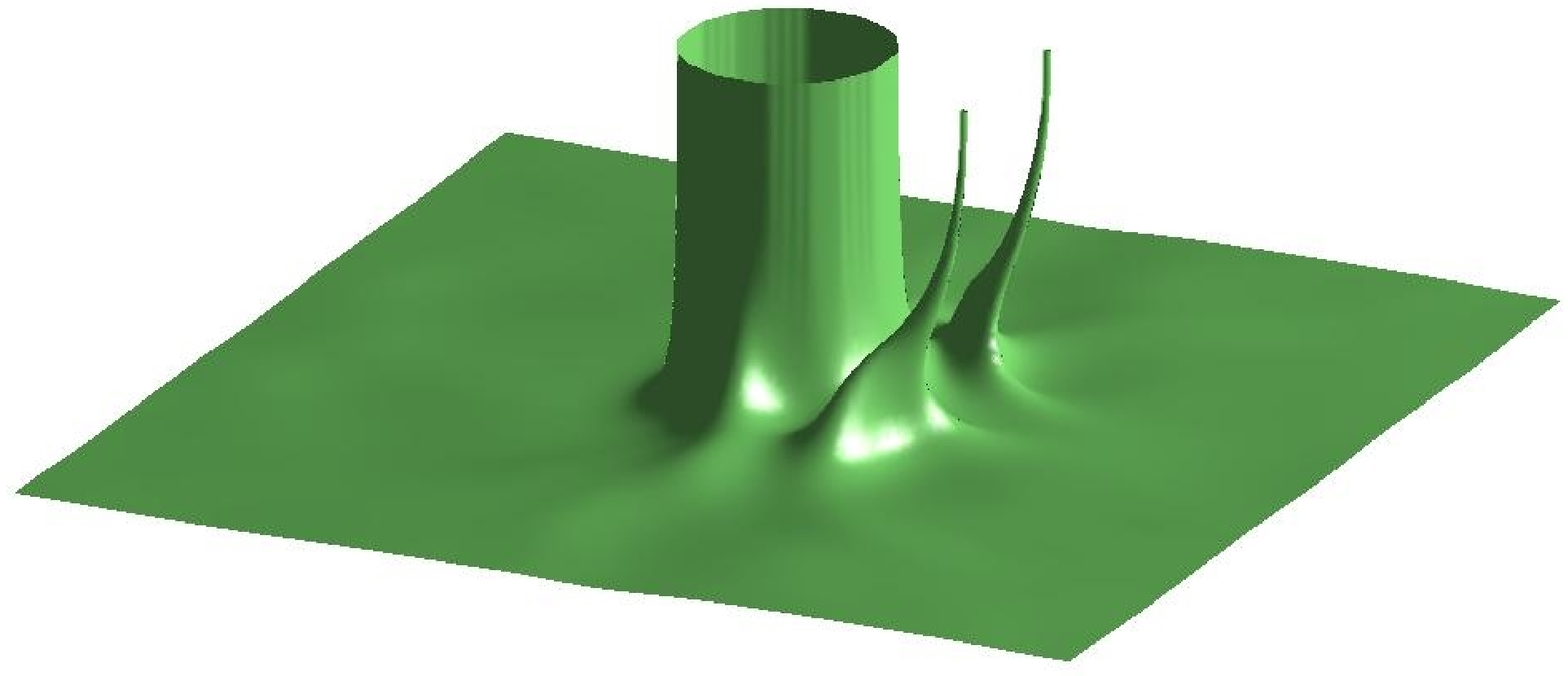}
$\;$
}
\centerline{
            \epsfxsize=2.6 in\epsfbox{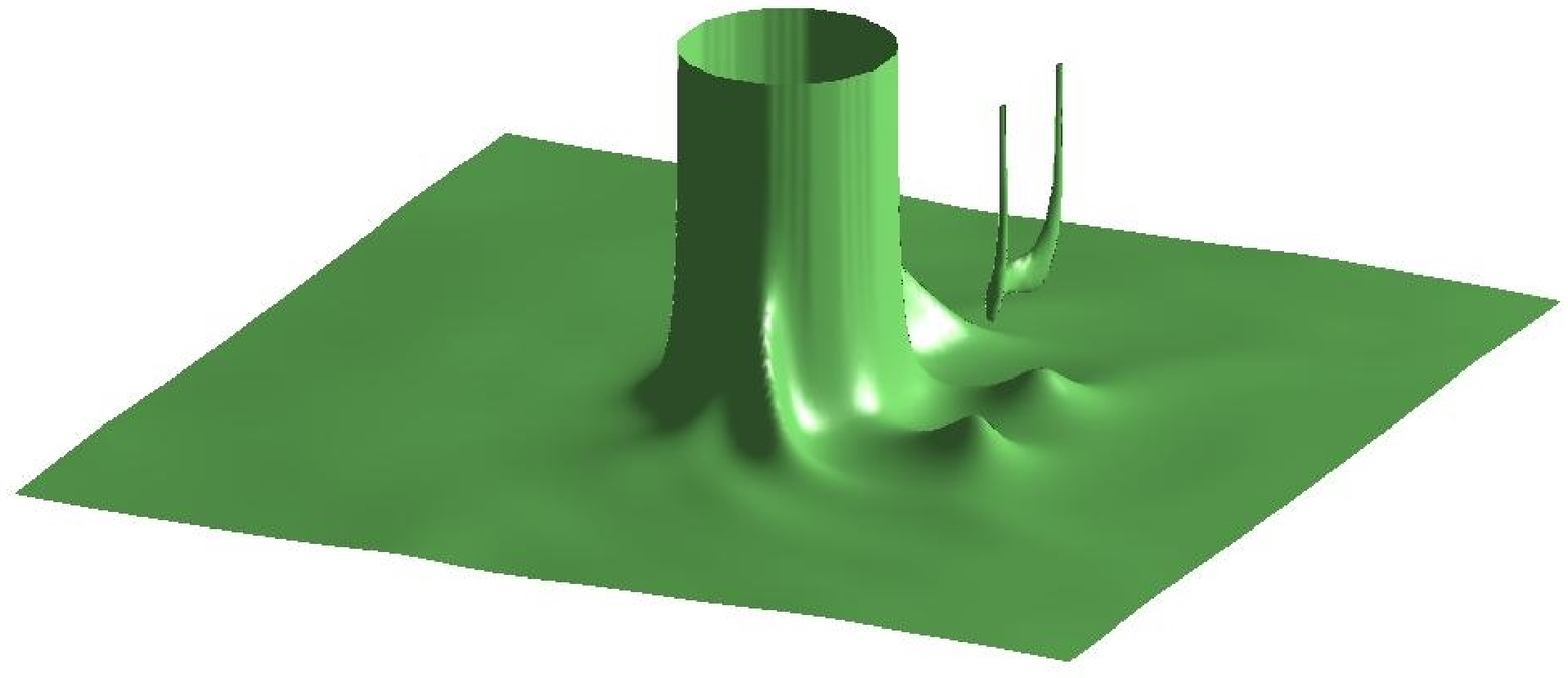}}
\caption{A sequence of isosurface snapshots of $|u|$ for $v=0.28$: 
 a) formation of vortex handles, b) detachment from obstacle, c)
bending of vortex tubes  and d) formation of vortex {\em half} 
rings.}
\vspace{-0.1in}
\label{fig4}
\end{figure}

 Let us describe the solutions for $v/c_s >0.25$ illustrated in Figure 
\ref{fig4}.
 The vortex handles seem to first nucleate near $z=0$
and are top connected to the obstacle. As time increases, the bottom
ends move  away from the obstacle in a slightly down stream 
direction while the top end moves up along the obstacle (Figure \ref{fig4}a).
 When the top ends of the vortices become close to  $z=L$, the
bottom ends reverse their trend of moving away from obstacle. Instead,
they  move back to the bottom of the obstacle, as if the handles
prefer certain curvature (Figure \ref{fig4}b).
Eventually, the top ends of the handle  move away from the obstacle and
produce a pair of vortex tubes with their bottom ends at the bottom of
the obstacle (Figure \ref{fig4}c). 
The handles merge into a half vortex ring, this half
ring moves both upward and downstream (Figure \ref{fig4}d). Near $z=0$, the solution can be approximated by the  
solution (\ref{sol}) and this solution  does not have vortices, so the instability
creates the vortex but the vortex moves away.
 Vortex detachment happens only at sufficiently high density, in the region where the nonlinear term in the equation dominates.  The direction of the vortex displacement is due to the velocity of the flow and the self interaction of the vortex on itself, which gives a movement along its normal vector.
Meanwhile, while the vortex ring starts to
detach from the obstacle,  another pair of vortex
handles is forming near the obstacle. 
The above process repeats itself.
 Note that we have truncated the domain close to the boundary of the cloud, so that the  half ring we compute would correspond to a closed  ring in the experiments.

 We have to point out that the critical velocity we have found for the onset of vortex shedding is lower than the critical velocity for the 2D problem at $z=L$. In this case $v_{2D}/c_s=0.35$. So the inhomogeneity in the condensate lowers the critical velocity from the 2D value.
 One can check that for different $L$, the critical velocity does not change.
This is verified by our numerical computation where 
we have used two boxes with one about 50\% higher in $z$
than the other, and there is little change in the
drag plots, nor there is any significant difference in the
dynamic behavior of the solutions.

In the experiments \cite{R,RO,O}, the drag is plotted vs velocity and a critical velocity can be defined when a sharp slope is observed in the drag plot.
 The critical velocity in \cite{R} is very similar to ours, though slightly smaller. This is certainly due to the finite extent of the condensate in the $x$, $y$ direction. Indeed, our simulations have not taken into account that the cloud is narrower in the $y$ direction than along the $x$. We can check that for the 2D problem, this geometry lowers the velocity. On the other hand, our computations indicate that the inhomogeneity in the $z$ direction and the soft boundary of the laser beam are well accounted for by our problem.


\hfill

{\bf Summary.}
We have studied the onset of dissipation in the Painlev\'e boundary layer of a BEC when  a 
detuned laser beam is moved in the condensate. We do a change of frame and
 blow up the low density region near the boundary of the cloud to write the equation for the wave function in this region: $z=0$ is now the boundary of the cloud and $z$ large is the center.  For small velocity, there is a drag around the obstacle due to radiation, but no vortex is generated: it is a stationary flow, which  is supersonic near $z=0$, but subsonic for $z$ larger. On the other hand, when the critical velocity is reached, the instability propagates towards the top, a vortex handle is nucleated and detaches from the obstacle to form vortex rings that move away. Our aim was to understand the origin of vortex shedding. The critical velocity is lower than for  the 2D problem.  There is a drag for all velocity, it increases smoothly with the velocity, and there a significant increase at the onset of vortex shedding.

\begin{acknowledgments}
\vspace{-0.1in}
 The authors would like to acknowledge discussions with
 Vincent Hakim and Marc Etienne Brachet. 
 Qiang Du is supported in part by a NSF grant DMS-0196522.
\end{acknowledgments}


\end{document}